\UseRawInputEncoding
\documentclass[prx,twocolumn,superscriptaddress,floatfix,nopacs]{revtex4}
\usepackage{graphicx,amsfonts,amssymb,amsmath,hyperref,enumerate,footnote}

\newif\ifhyper
\hypertrue
\ifhyper
\hypersetup{
   citecolor = {green},
   colorlinks = {true}, 
   urlcolor = {blue} 
}
\fi

\newcommand{\beq}{\begin{equation}}
\newcommand{\eeq}{\end{equation}}
\newcommand{\beqa}{\begin{eqnarray}}
\newcommand{\eeqa}{\end{eqnarray}}
\newcommand{\ket} [1] {\vert #1 \rangle}
\newcommand{\bra} [1] {\langle #1 \vert}

\def\bra#1{\langle#1\vert}
\def\ket#1{\vert#1\rangle}

\def\Longarrow{\protect\@lra}
\def\@lra{\relbar\joinrel\relbar\joinrel\relbar\joinrel%
          \relbar\joinrel\rightarrow}

\begin{document}

\title{Variational Quantum Continuous Optimization: \\ a Cornerstone of Quantum Mathematical Analysis}

\author{Pablo Bermejo}
\affiliation{Multiverse Computing, Paseo de Miram\'on 170, E-20014 San Sebasti\'an, Spain}
\affiliation{Donostia International Physics Center, Paseo Manuel de Lardizabal 4, E-20018 San Sebasti\'an, Spain}

\author{Rom\'an Or\'us}
\affiliation{Multiverse Computing, Paseo de Miram\'on 170, E-20014 San Sebasti\'an, Spain}
\affiliation{Donostia International Physics Center, Paseo Manuel de Lardizabal 4, E-20018 San Sebasti\'an, Spain}
\affiliation{Ikerbasque Foundation for Science, Maria Diaz de Haro 3, E-48013 Bilbao, Spain}

\begin{abstract} 

Here we show how universal quantum computers based on the quantum circuit model can handle mathematical analysis calculations for functions with continuous domains, without any digitalization, and with remarkably few qubits. The basic building block of our approach is a variational quantum circuit where each qubit encodes up to three continuous variables (two angles and one radious in the Bloch sphere). By combining this encoding with quantum state tomography, a variational quantum circuit of $n$ qubits can optimize functions of up to $3n$ continuous variables in an analog way. We then explain how this quantum algorithm for continuous optimization is at the basis of a whole toolbox for mathematical analysis on quantum computers. For instance, we show how to use it to compute arbitrary series expansions such as, e.g., Fourier (harmonic) decompositions. In turn, Fourier analysis allows us to implement essentially any task related to function calculus, including the evaluation of multidimensional definite integrals, solving (systems of) differential equations, and more. To prove the validity of our approach, we provide benchmarking calculations for many of these use-cases implemented on a quantum computer simulator. The advantages with respect to classical algorithms for mathematical analysis, as well as perspectives and possible extensions, are also discussed.  

\end{abstract}

\maketitle

\section{Introduction}

The field of quantum computing is evolving rapidly. What looked like science-fiction a few years ago, is now starting to become a reality. Thanks to recent experimental breakthroughs, we now can enjoy the first commercial quantum processors. Such machines are noisy and have a limited number of qubits (i.e., they are Noisy Intermediate-Scale Quantum - NISQ - devices \cite{nisq}), and are thus far from ideal. Following the trends in history, we should expect to have more powerful and noise-resistant processors in the future. But, as of today, our quantum computers are made of a small number of noisy qubits.   

Now, here comes a hard question: can we do something useful with these imperfect quantum machines? The qualified answer to this question is \emph{yes}. To put it in perspective, think of the evolution of classical computers. One would never use a computer from the 1930's to develop, say,  a speech recognition system. But those machines were definitely useful at cracking some cryptographic codes. The situation with current quantum computers is, in a way, similar to that of classical computers in the 30's. Of course we cannot hope to use today's quantum processors for very complex tasks such as, say, factoring a 617-digit number (as used in the RSA cybersecurity protocol). But they are certainly useful for other tasks. And of course, what consumes mental energy is to find those tasks, where today's quantum computers can begin to offer real, practical value. So, what can we already do with NISQ processors that is actually useful? 

A good example where NISQ devices can start offering value is \emph{optimization}. By developing hybrid quantum-classical solutions, present-day quantum annealers can handle hard optimization problems in real scenarios \cite{qopt1, qopt2, qopt3, qopt4, qopt5, qopt6, qopt7, qopt8, qopt9}. Another good example is that of \emph{machine learning}, where few-qubit universal quantum systems are, at least, as powerful as deep neural networks for supervised and unsupervised learning \cite{latorre, clustering}, thanks to the use of variational quantum optimization methods \cite{VQE, QAOA}. 

In this paper we follow the above philosophy, and show how few-qubit quantum computers can already implement arbitrary multi-dimensional function calculus in a remarkably efficient way. The basic building block of our approach is a variational quantum algorithm to optimize functions with continuous domains. In our approach, each qubit in the quantum circuit encodes up to three continuous variables in the parameters of the Bloch sphere: two angles and one radius. By combining this with single-qubit quantum tomography, a variational optimization of the circuit parameters allows to find the extreme values of multidimensional functions in a purely-analog and fast way. 

Based on the above, we then show how our quantum optimization algorithm can be used to implement, essentially, any type of calculus on functions with continuous domains. As a first step, we show how to compute arbitrary series expansions of any function, and in particular Fourier (harmonic) analysis, with the coefficients of the expansions calculated via continuous quantum optimization. And, then, with this tool at our disposal, we explain how this allows us to do pretty much any calculation, including definite integrals of multidimensional functions, solving (systems of) differential equations, and more. 

The quantum optimization algorithm presented here is the cornerstone of an efficient quantum toolbox for mathematical analysis on NISQ processors. To put it in perspective, a quantum computer of 127 qubits, such as the IBM-Q System One \cite{ibm}, could use our algorithms to analyze 381-dimensional functions directly in the continuum. Additionally, we also find that the classical simulation of these quantum algorithms is \emph{very fast}, comparable in performance to standard classical computational software. 

This paper is organized as follows: in Sec.\ref{sec2} we present the building block of our construction, namely, our quantum algorithm for continuous optimization. In Sec.\ref{sec3} we explain different applications of this algorithm, including series expansions, Fourier analysis, definite integral calculations, and differential equations, where we provide also benchmarks and examples of such calculations  via simulation of a quantum computer. Finally, Sec.\ref{sec4} discusses the efficiency of our methods, and wraps up with our conclusions and perspectives for future work. 

\section{Continuous quantum optimization}
\label{sec2}

Let us start by presenting our quantum algorithm for optimizing multidimensional functions with continuous domains. In the following, we discuss the basics of the problem and the proposed quantum algorithm, which is based on a variational quantum circuit with a peculiar encoding of the function's variables. 

\subsection{The problem} 

Without loss of generality, let us consider a real scalar function $f(\vec{x})$ of $m$ real variables $\vec{x} = (x_1, x_2, \cdots, x_m)$. The domain of each variable $x_\alpha$ is $D_\alpha \subseteq {\mathbb R}$ for all $\alpha = 1, 2, \cdots, m$. All the derivations that we explain in this paper can be directly translated to more complex scenarios such as, e.g., real and complex vector and tensor fields of real and complex variables. 

The statement of the problem is simple: find the minimum of  $f(\vec{x})$, i.e.,  
\beq
\min_{\vec{x}} f(\vec{x}). 
\eeq
As opposed to a discrete optimization problem, which is the usual case addressed by quantum computing, here the variables used in the objective function are continuous. At this stage, we do not demand more constraints on the function or the domains. In typical applications, though, one may focus on, e.g., continuous and differentiable functions. From the use-case perspective, continuous optimization is at the very core of many real-world problems in mathematical science and engineering, including the design of biomolecules, financial portfolio optimization, and fluid dynamics, among many others. 

\subsection{Variable encoding} 

We now explain one of the main characteristics of our quantum algorithm: the way we encode the continuous variables $x_\alpha$ in the degrees of freedom of a quantum computer. 

\begin{figure}
	\centering
	\includegraphics[width=0.6\linewidth]{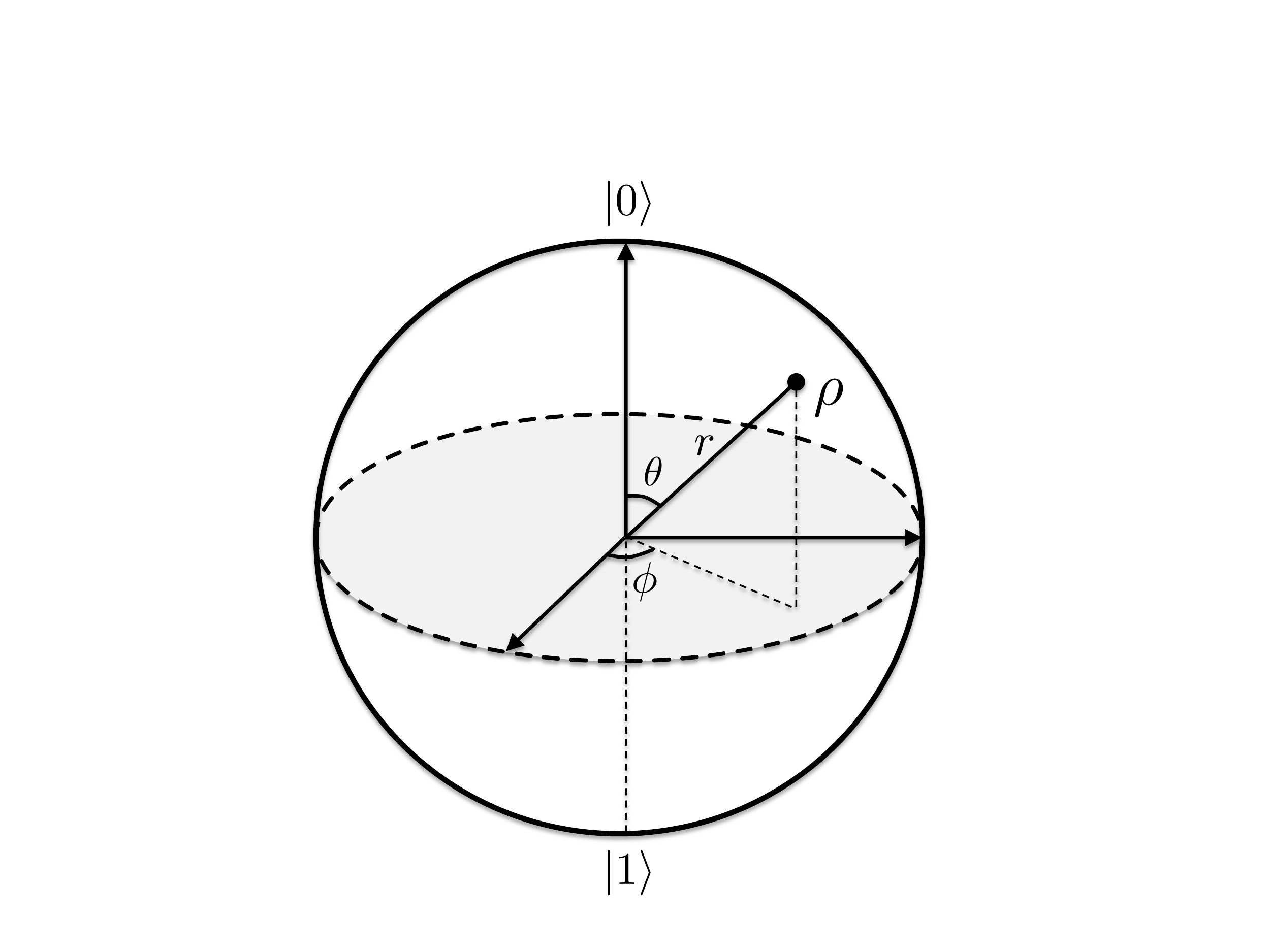}
	\caption{Bloch sphere of one qubit. The mixed-state state $\rho$ is a point in the interior of the sphere and is characterized by three continuous parameters $(\theta, \phi, r)$ as specified in the text. Pure states correspond to points in the surface of the sphere, so that $r=1$, and depend on two angular variables $(\theta, \phi)$.}
	\label{fig1}
\end{figure}

To see the encoding in perspective, let us remind the basic degrees of freedom of a single qubit. In general, a qubit is described by a point of a Bloch sphere, see Fig.\ref{fig1}. Pure states of a qubit correspond to points on the surface of the sphere, and can be parametrized by two angles, which we call $\theta \in [0, \pi]$ and $\phi \in [0, 2\pi)$. Accordingly, the pure state $\ket{\psi}$ of a qubit is given by the well-known formula 
\beq
\ket{\psi} = \cos\left( \frac{\theta}{2} \right) \ket{0} + e^{i \phi} \sin\left( \frac{\theta}{2} \right) \ket{1},
\eeq
with $\{ \ket{0}, \ket{1} \}$ the orthogonal computational basis of the Hilbert space. Similarly, mixed states of a qubit correspond to points in the interior of the Bloch sphere. These states correspond to the single-qubit configurations where the qubit is entangled with some other quantum system, and therefore the qubit's reduced density matrix is no longer a pure state. Mathematically, the mixed state $\rho$ of a qubit is written as 
\beq
\rho = \frac{1}{2} \left( {\mathbb I} + \vec{n}\cdot \vec{\sigma} \right), 
\eeq
where ${\mathbb I}$ is the $2 \times 2$ identity matrix, $\vec{n} = (n_x, n_y, n_z)$ is a three-dimensional vector of real components, and $\vec{\sigma} = (\sigma_x, \sigma_y, \sigma_z)$ is a three-dimensional vector of the three Pauli matrices $\sigma_x, \sigma_y$ and $\sigma_z$. In this notation, the qubit is in a pure state (i.e., in the surface of the Bloch's sphere) iff $| \vec{n} | = 1$, in which case the vector is given by $\vec{n} = (\sin \theta \cos \phi, \sin \theta \sin \phi, \cos \theta )$ in spherical coordinates, so that $\rho = \ket{\psi} \bra{\psi}$. In addition, the qubit is in a mixed state whenever $|\vec{n}| < 1$, so that $\vec{n} = r (\sin \theta \cos \phi, \sin \theta \sin \phi, \cos \theta )$, with $r < 1$ being the radial coordinate of a point inside the sphere. 

The main point of our encoding is to use the continuous variables of a qubit to codify the function's variables $x_\alpha$. In this way, for a generic qubit mixed state we can encode three variables, i.e., 
\beqa
[ x_\alpha, x_\beta, x_\gamma ] &\rightarrow& [\theta, \phi, r], \nonumber \\
D_\alpha \rightarrow [0, \pi], ~~ D_\beta &\rightarrow& [0, 2 \pi), ~~ D_\gamma \rightarrow [0,1), 
\eeqa
where we map function variables to the continuous qubit variables, with a corresponding embedding of the domains $D_\alpha, D_\beta, D_\gamma$. In this way, we can encode up to three continuous variables per mixed-state qubit. Similarly, if the qubit is in a pure state, then $r = 1$ and we can encode up to two continuous variables per pure-state qubit.  

\subsection{Optimization Algorithm}

We now turn into solving the problem of continuous optimization, using a quantum computer in combination with the encoding that we just introduced. For this, we will make use of a variational quantum circuit to minimize the function $f(\vec{x})$, following the procedure  of a variational quantum eigensolver (VQE) \cite{VQE} but with our embedding of the variables. 

\begin{figure}
	\centering
	\includegraphics[width=0.9\linewidth]{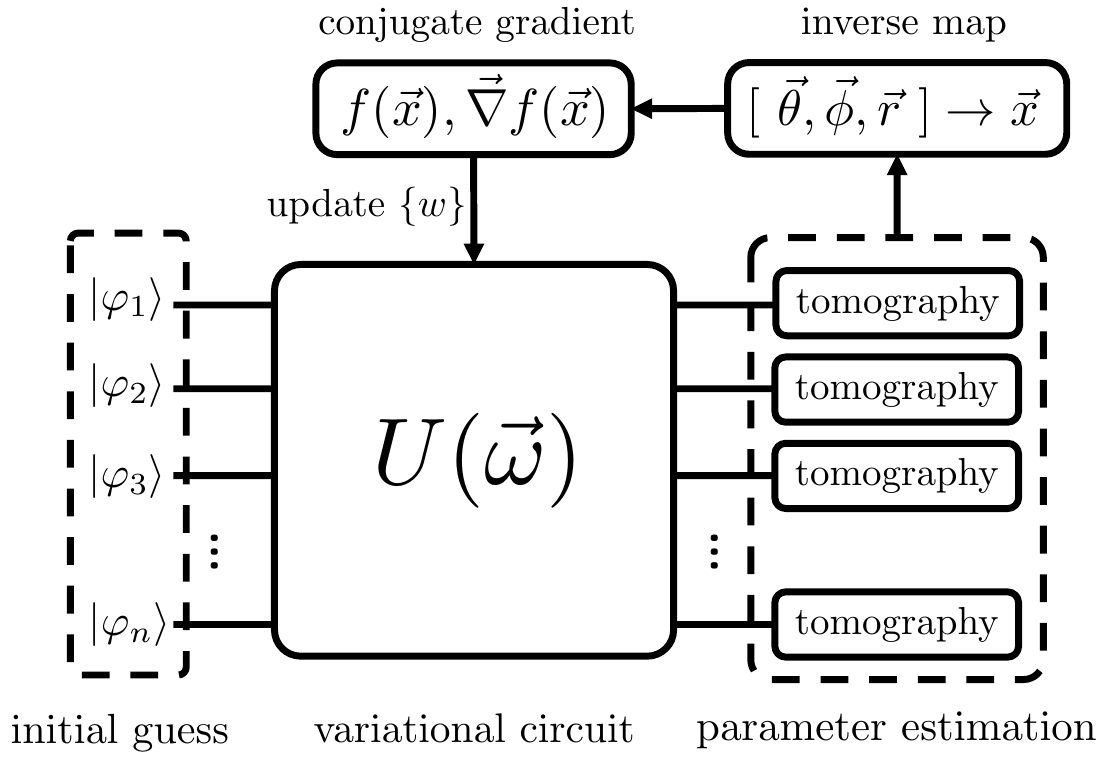}
	\caption{Sketch of the variational continuous quantum optimization algorithm, as discussed in the text. We implement a variational quantum circuit $U(\vec{\omega})$ with a set of variational parameters $\vec{\omega}$. Afterwards, we estimate the qubit parameters $[ \vec{\theta}, \vec{\phi}, \vec{r}]$ for all qubits via single-qubit tomography, and from these the value of the function variables $\vec{x}$. This allows us to estimate the function $f(\vec{x})$ as well as its gradient $\vec{\nabla} f(\vec{x})$, and then update the circuit parameters $\vec{\omega}$ via, e.g., conjugate gradient. See the text for an in-depth discussion.} 
	\label{fig2}
\end{figure}
 
The main idea of our approach is schematically represented in Fig.\ref{fig2}. We implement a variational quantum circuit, where different gates depend on parameters that are to be optimized in order to minimize the value of the cost function $f(\vec{x})$. In this setting, the circuit can be as simple or as complex as we wish, exactly in the same way as any other VQE calculation. The main point of our approach, though, is that we are encoding the variables  of the function in the  degrees of freedom of the individual qubits. Thanks to this, if we consider qubit mixed states, we can then use a variational quantum circuit of $n$ qubits to optimize functions of up to $m = 3n$ variables. Similarly, if we consider qubit pure states, we can then optimize functions up to $2n$ variables. The overall idea of the algorithm is as follows: 
\begin{enumerate} 
\item{Initialize the quantum circuit parameters to some initial guess.}
\item{Run the circuit over $n$ qubits.}
\item{Implement quantum state tomography of the individual qubits  to estimate parameters $[\theta, \phi, r]$ for each mixed-state qubit, or $[\theta, \phi]$ for each pure-state qubit.}
\item{Estimate the vector of variables $\vec{x}$ from the qubit parameters.}
\item{Estimate the value of the function $f(\vec{x})$ and its gradient $\vec{\nabla} f (\vec{x})$ at that point $\vec{x}$.}
\item{Update the circuit parameters via, e.g., gradient descent, and iterate until convergence.}
\end{enumerate}
This is certainly a basic version of an optimization algorithm based on gradient descent and a variational quantum circuit. The whole point of our approach is to encode the function's variables into the qubits' variables, and determine these via single-qubit quantum state tomography. Importantly, what we are determining here are continuous variables, i.e., continuous states of individual qubits, and we are never relying on a discrete set of individual qubit states. The optimization is, thus, purely analog. 

The above basic algorithm can be refined in many ways. One could,  for instance, implement different algorithms rather than gradient descent, and even gradient-free methods for those functions with a chaotic gradient. Additionally, we could also implement an encoding of the variables, not on the single-qubit parameters, but rather on the continuous parameters of an entangled many-qubit wavefunction. This would offer a greater representation power, though at the expense of being more complex to read out via multiqubit quantum state tomography. From a practical perspective also, notice that encoding the variables in the parameters of mixed state qubits means that the quantum state produced by the variational quantum circuit is a multi-qubit entangled state, whereas the encoding in pure state qubits produces a separable non-entangled state. As an additional refinement, one could also consider taking entangled states as initial guesses instead of product states. In this respect, we could use an educated Tensor Network (TN) \cite{TN, TN2, TN3, TN4} efficiently prepearable on a quantum computer (such as a Matrix Product State) as initial guess, and then let the variational quantum circuit add/remove entanglement on top of it. Such initial TN state could be, say, the outcome, or even an intermediate-step state, of a TN optimization algorithm. 

\subsection{Some examples}Ê

As a first example, let us consider a system of 7 qubits, where we encode 2 variables per pure-state qubit. We then implement a variational quantum circuit to optimize the 14-variable trigonometric function 
\beq
f(\vec{x}) = \sum_{\alpha = 1}^5 \sin(x_\alpha) + \sum_{\alpha = 6}^{10} \cos(x_\alpha) + 4 \sum_{\alpha = 11}^{14} \left(\cos(x_\alpha)\right)^2, 
\eeq
with $x_\alpha \in [0, 2 \pi)$ for all $\alpha$. In our simulations we see that even a simple variational quantum circuit made of 1-qubit rotations and sequential 2-body CNOT gates can find extreme configurations such that $f(\vec{x}) \approx 0$ remarkably fast, al most with arbitrary precision. What is more, we observe that the algorithm is very stable, in the sense that once an extreme has been found, the variational quantum circuit tends to stay there and not jump to a different configuration. Without loss of generality, we'll stick to this type of quantum circuit for the rest of the paper. 

Next, we considered the optimization of the 4-variable function 
\beq
f(x_1, x_2, x_3, x_4) = \sin\left(\frac{x_1}{x_4 \cos \left( \log\left( x_1^2 x_2 x_3^{-1} \right)\right)}\right). 
\eeq
In less than three seconds, the simulation of the quantum algorithm with 2 pure-state qubits was able to find a extreme configuration, with $x'_1 = 3.201, x'_2 = 0.396, x'_3 = 3.729, x'_4 = 0.647$ and $f(x'_1, x'_2, x'_3, x'_4) \approx - 0.999$. 

In order to push the algorithm further, we then considered the minimization of the 28-variable function 
\beqa
&&f(\vec{x}) = \sin\bigg(  x_1 x_2^{-1} \cos\left(\log\left( \frac{x_1^3 x_3}{x_4} \right) \right)\sin\left(\frac{x_5}{x_2}\right) \nonumber \\ 
&+& \cos\left( x_6^{1/2}x_1 x_5^{-2} \right)  - x_9^2 \left( x_{10} - x_{11} x_1 x_4^{-1} \right) \nonumber \\ 
&+& \sin\left(\frac{x_7^3}{x_1x_3 + x_4}\right)\cos\left( x_8x_3^{-1} \sin(x_7) \right) \nonumber \\
&+& \cos\left(x_{12}^2 - x_9 x_{10} \right) + \cos\left( \frac{x_{21} x_{22}}{x_{23}} - \sin(x_{24}) \right) \nonumber \\ 
&+& \cos(x_{13}x_{14}) \log \left( \frac{x_{15}}{x_{16}} + x_{14} x_{15}^2 \sin\left( x_{13} \cos\left( \frac{x_{16}}{x_{15}} \right)\right)\right) \nonumber \\ 
&+& \sin\left( \frac{x_1^2 x_{17}}{x_{18}} + \cos\left( \cos\left( \frac{x_{19}}{x_{20}} \right)\right) \right) \nonumber \\ 
&+&  \sin\left( x_{25} x_1 x_6^{1/2} x_{26} \right) + \cos\left( x_{27} x_{28}^{2} \right) \nonumber \\\ 
&-& x_3 \log\left( \left( \frac{x_{27}x_{28}}{x_{21}} - \sin(x_5 x_{11}) \right) \right) \bigg). 
\eeqa
Using the same structure of variational quantum circuit as for the previous examples, for 14 qubits encoding in the variables in pure states, the simulation of the quantum algorithm was able to find configurations such that $f(\vec{x}) \approx - 0.999$ in just a few seconds. 

\bigskip 

The algorithm described and benchmarked here allows to optimize functions of continuous variables, directly in an analog way, using a finite number of qubits. In the following we show how this can be used as the building block of a toolbox for mathematical analysis of functions with gate-based quantum computers. 

\section{Mathematical Applications}
\label{sec3}

The implications of our quantum optimization algorithm go well beyond optimization itself. In fact, optiization is just the tip of the iceberg. In this section we elaborate the details of how this algorithm can be used to solve a wide variety of problems in mathematical analysis using a quantum computer. 

\subsection{Series expansions}
 
 A series expansion is an expansion of a function into a series or infinite sum. As is well-known in science and engineering, series expansions are an ideal way to express and approximate functions via truncation of the series, allowing to compute and approximate further, more complex calculations. Good examples of this are Fourier series in mathematics and engineering, as well as Taylor expansions, at the basis of, e.g., perturbation theory in physics. The number of examples and applications is unending. 
 
Let us consider, again without loss of generality, a scalar function $f(\vec{x})$ of $m$ real variables, $x_\alpha \in {\mathbb R}$ for $\alpha = 1, 2, \cdots, m$. A generic series expansion is given by
\beq
f(\vec{x}) = \sum_{l = 0}^\infty c_l g_l(\vec{x}), 
\eeq
where functions $g_l(\vec{x})$ are a orthonormal basis of the vector space of functions in the domain $D$ of $\vec{x}$ with respect to a weight function $w(\vec{x})$, i.e., 
\beq
\langle g_l, g_{l'} \rangle \equiv \int_{D} w(\vec{x}) g^*_l(\vec{x}) g_{l'}(\vec{x}) d \vec{x} = \delta_{l l'}, 
\label{prod}
\eeq
and coefficients $c_l$ are given by the scalar product 
\beq 
c_l \equiv  \int_{D} w(\vec{x}) g^*_l(\vec{x})f(\vec{x}) d \vec{x} . 
\eeq
As is well known, one may also consider approximations to the function $f(\vec{x})$ by truncating such series, i.e., 
\beq
f(\vec{x}) \approx \sum_{l = 0}^{K-1} c_l g_l(\vec{x}), 
\eeq
keeping the first $K$ terms in the series, and where the error is usually quantified with upper bounds. A more refined approximation to this one, is to actually compute the \emph{optimal} coefficients when involving only $K$ terms in the expansion. This is the solution to the optimization problem 
\beq
\min_{\vec{c}} \left(\int_D w(\vec{x}) \left|f(\vec{x}) - \sum_{l = 0}^{K-1} c_l g_l(\vec{x})\right|^2 d\vec{x} \right)^{1/2},     
\eeq 
where we are minimizing the quadratic distance with respect to the product in Eq.(\ref{prod}), and where we define $\vec{c} \equiv (c_0, c_1, \cdots, c_{K-1})$ as a vector of coefficients. 

Given that computing the above integral may be cumbersome, we can approximate it with good accuracy by discretizing it using the Riemann sum 
\beq
\min_{\vec{c}} \left(\sum_{\mu = 1}^{M} w(\vec{x}_{\mu}) \left| f(\vec{x}_{\mu}) - \sum_{i = 0}^{K-1} c_l g_l(\vec{x}_{\mu})\right|^2 \Delta\vec{x}_\mu \right)^{1/2},     
\eeq 
where $\vec{x}_\mu$ are $M$ points in the domain $D$. The solution to this problem is equivalent to optimizing the squared distance 
\beqa
\min_{\vec{c}} \left(\sum_{\mu = 1}^{M} w(\vec{x}_{\mu}) \left|f(\vec{x}_{\mu}) - \sum_{i = 0}^{K-1} c_l g_l(\vec{x}_{\mu})\right|^2 \Delta\vec{x}_\mu \right), 
\label{suma}
\eeqa
which can be rewritten as 
\beqa
\min_{\vec{c}^\dagger, \vec{c}} \sum_{l,l' = 0}^{K-1} A_{ll'} c^*_l c_{l'} + \sum_l c^*_l b_l + b^*_l c_l + q, 
\eeqa 
where matrix $A_{ll'}$, vector $b_l$ and scalar $q$ can be easily computed from Eq.(\ref{suma}). In compact notation we have 
\beqa
\min_{\vec{c}^\dagger, \vec{c}} \left(\vec{c}^{\dagger} A \vec{c} + \vec{c}^\dagger \cdot \vec{b}  + \vec{b}^\dagger \cdot \vec{c} + q\right) = \min_{\vec{c}^\dagger, \vec{c}} F(\vec{c}^\dagger, \vec{c}).  
\label{opto}
\eeqa 
The minimum of this quadratic form can be computed by evaluating the gradient with respect to, e.g., $\vec{c}^\dagger$, and making it equal to zero, i.e., 
\beq
\frac{d}{d \vec{c}^\dagger} F(\vec{c}^\dagger, \vec{c}) = A \vec{c} + \vec{b} = 0, 
\eeq
which provides the closed-form solution 
\beq
\vec{c} = A^{-1} \vec{b}. 
\eeq

While the above solution is analytically valid, it may still be problematic to compute if matrix $A$ is ill-conditioned, so that the calculation of its inverse is numerically imprecise. Additionally, we may use other distance measures that are not necessarily quadratic, such as $l_p$-norms, and therefore not having any closed analytical solution. In such cases, it pays off to solve directly the optimization problem in Eq.(\ref{opto}). The coefficients of the optimal truncated expansion are then better computed directly by solving the corresponding optimization problem for the coefficients $c_i$. Such coefficients can be computed straightforwardly using the quantum optimization algorithm presented in the previous section. 

\begin{figure}
	\centering
	\includegraphics[width=0.9\linewidth]{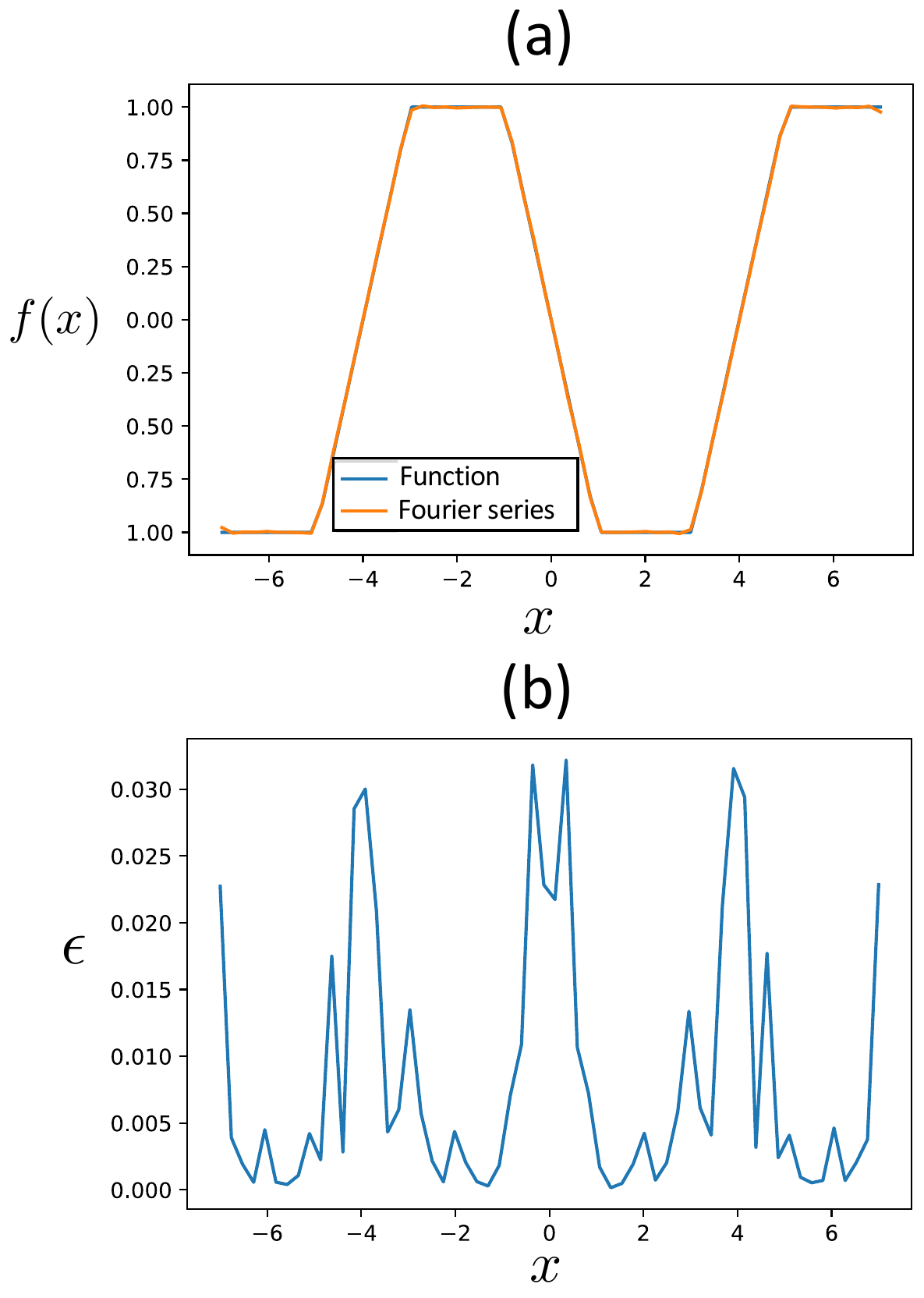}
	\caption{[Color online] (a) Step function $f(x)$ and Fourier approximation (almost indistinguishable); (b) relative error between exact and Fourier approximation.}
	\label{fig3}
\end{figure}

\subsection{Fourier analysis} 

The case of Fourier analysis is a particular example of the above. For instance, for a one-variable function of period $2L$ we would have the expression 
\beq
f(x) = \sum_{l = -\infty}^{\infty} c_l e^{i 2 \pi l x / L} \approx \sum_{l = -K/2}^{K/2} c_l e^{i 2 \pi l x / L}, 
\eeq
with the last approximation involving $K$ Fourier modes. As a reminder, if the function is not periodic, then $2L$ is the size of the domain $D$ where we wish to find the expansion. Similarly, one could choose a Fouruier expansion in terms of trigonometric functions, such as 
\beq
f(x) \approx a_0 + \sum_{l = 1}^{K} \left(a_l \cos(\pi l x / L) + b_l \sin(\pi l x / L) \right). 
\eeq
Following the procedure explained in the previous section, the coefficients $c_l, a_l$ and $b_l$ can be found straightforwardly by solving a continuous optimization problem. This can be done using our continuous quantum optimization algorithm, allowing then to implement Fourier analysis directly in an analog way on a quantum computer. The procedure can be extended to multidimensional functions with similar effort. 

In order to benchmark the performance of this method, we have computed the Fourier series for a variety of functions. In Fig.\ref{fig3}(a) we show the approximation of a step function using 6 qubits, 16 Fourier modes, and a mesh of 80 points for the optimization, with the relative error in Fig.\ref{fig3}(b). We can see that the exact function and the approximation are indistinguishable with the naked eye, with a small relative error around $10^{-2} - 10^{-3}$. The calculation took just a few minutes on our laptops. 

\begin{figure}
	\centering
	\includegraphics[width=0.9\linewidth]{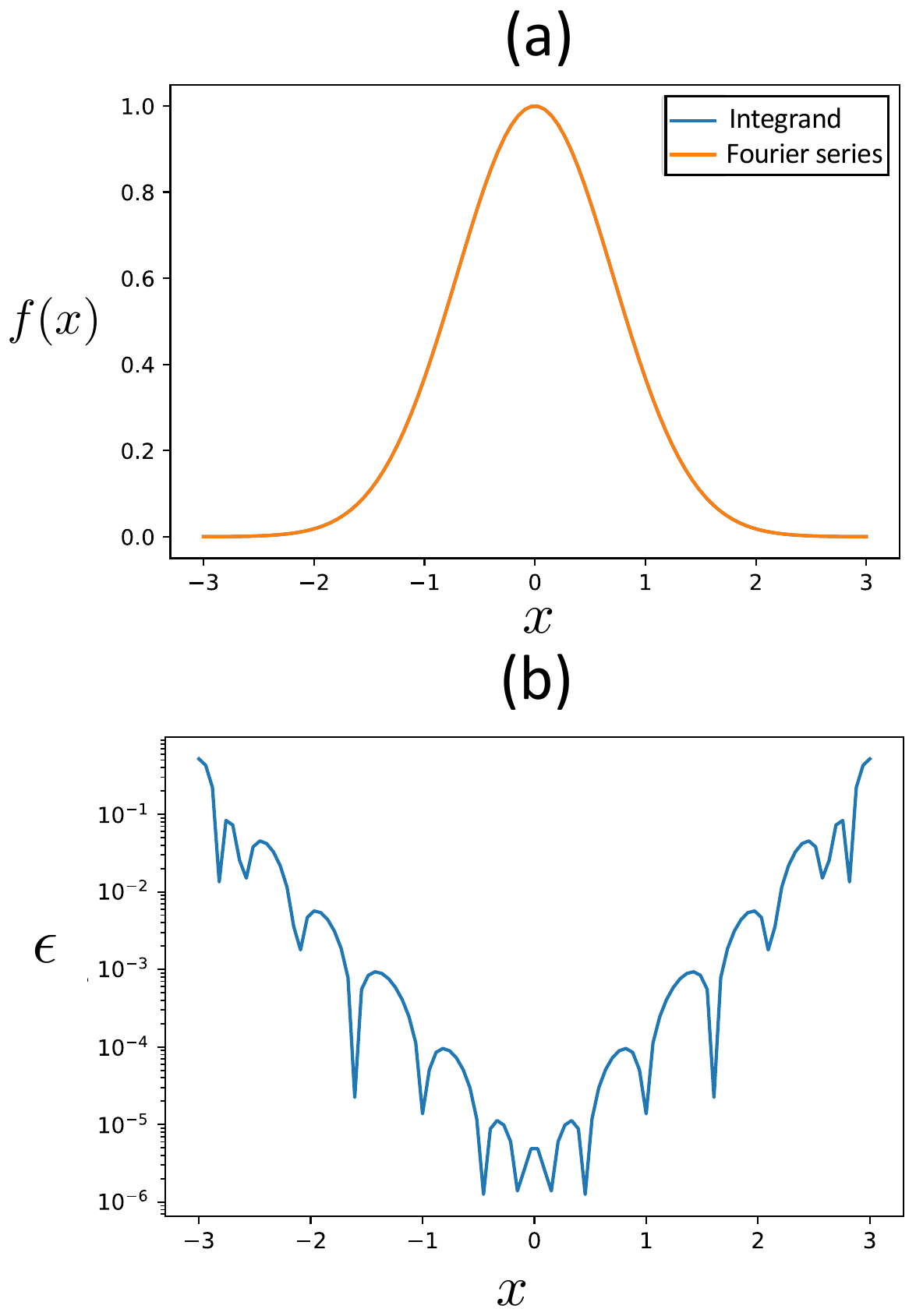}
\caption{[Color online] (a) Gaussian function $f(x) = e^{-x^2}$ and Fourier approximation (almost indistinguishable); (b) relative error between exact and Fourier approximation.}
	\label{fig4}
\end{figure}

\begin{figure}
	\centering
	\includegraphics[width=0.9\linewidth]{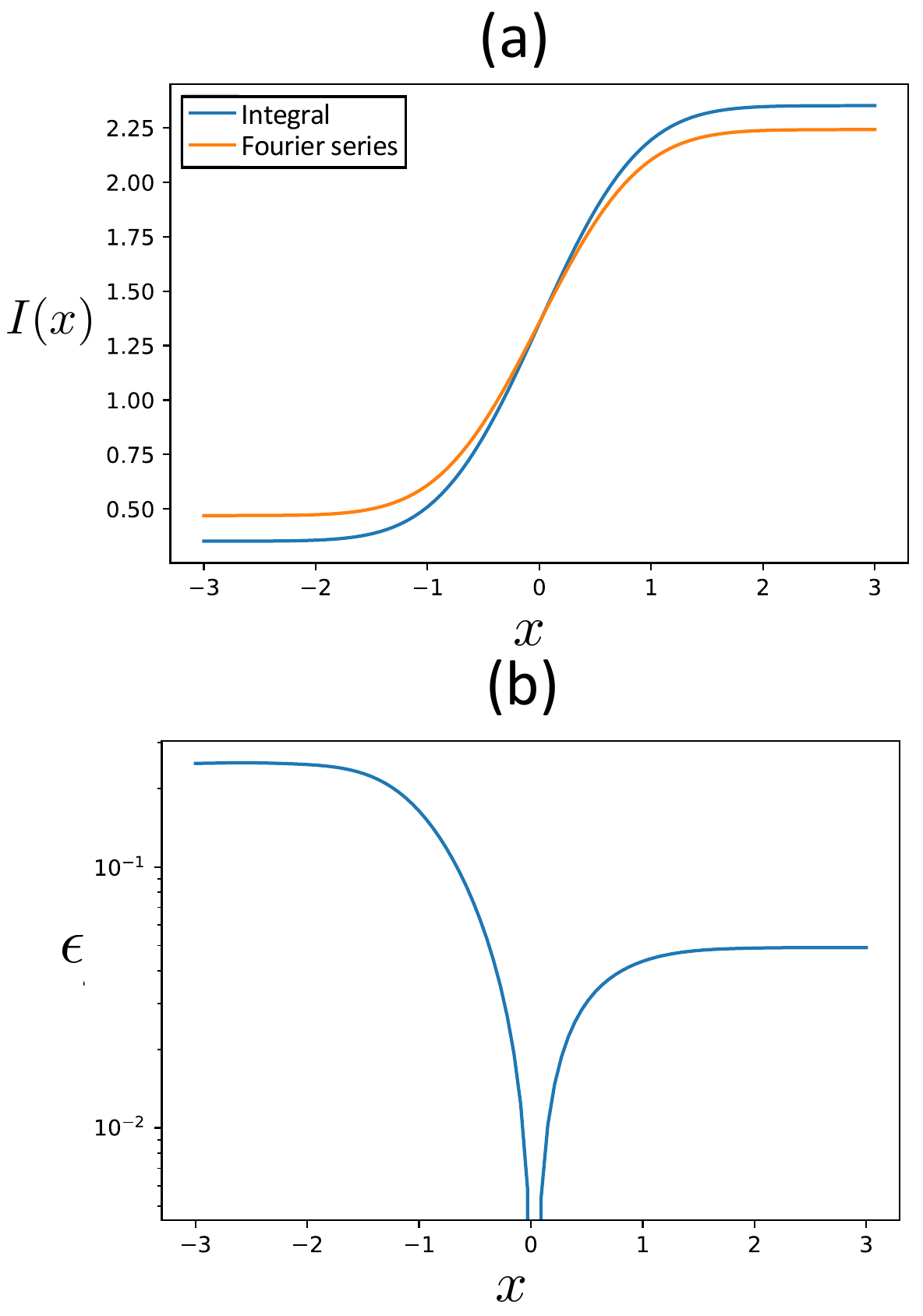}
	\caption{[Color online] (a) Integral of Fig.\ref{fig4} (error function), exact and Fourier approximation; (b) relative error between exact and Fourier approximation.}
	\label{fig5}
\end{figure}

\subsection{Integrals}
The calculation of complicated integrals is straightforward in our setting. For this, we can use Fourier analysis, so that the integral becomes remarkably easy to implement. Imagine, for instance, that we wish to compute the integral 
\beq
I(a,b) = \int_a^b f(x) dx, 
\eeq
where, without loss of generality, we focus for simplicity again on the case of a one-dimensional scalar function. Approximating it by a truncated Fourier series of exponentials in the interval $[a, b]$ we find  
\beqa
I(a,b) &=& \int_a^b f(x) dx \approx \sum_{l = -K/2}^{K/2} c_l \int_a^b e^{i 2 \pi l x /(b-a)} dx \\
&=& \sum_{l = -K/2}^{K/2} \frac{i c_l (b-a)}{2 \pi l} \left( e^{i 2 \pi l a /(b-a)} - e^{i 2 \pi l b /(b-a)} \nonumber \right), 
\eeqa
so that we have a closed formula for approximating the integral, once we have the coefficients $c_l$. 

As a practical example, we have computed the definite integral of a gaussian, i.e., the error function 
\beq 
I(a,b) = \int_a^b e^{-x^2} dx. 
\eeq
In Fig.\ref{fig4} we show the integrand and its Fourier approximation, as well as the relative error, using 8 qubits and 22 Fourier modes with an optimizatino mesh of 100 points. As we can see, the approximation is really good in few minutes, even with small error at the tails of the distribution where its value becomes exponentially small. In Fig.\ref{fig5} we show the integral and its Fourier approximation, also with the relative error, with the same numner of qubits and modes, and a mesh of 80 points for the optimization. The approximation is also really good, considering the exponentially-small contributions from the tails. 

\begin{figure}
	\centering
	\includegraphics[width=0.9\linewidth]{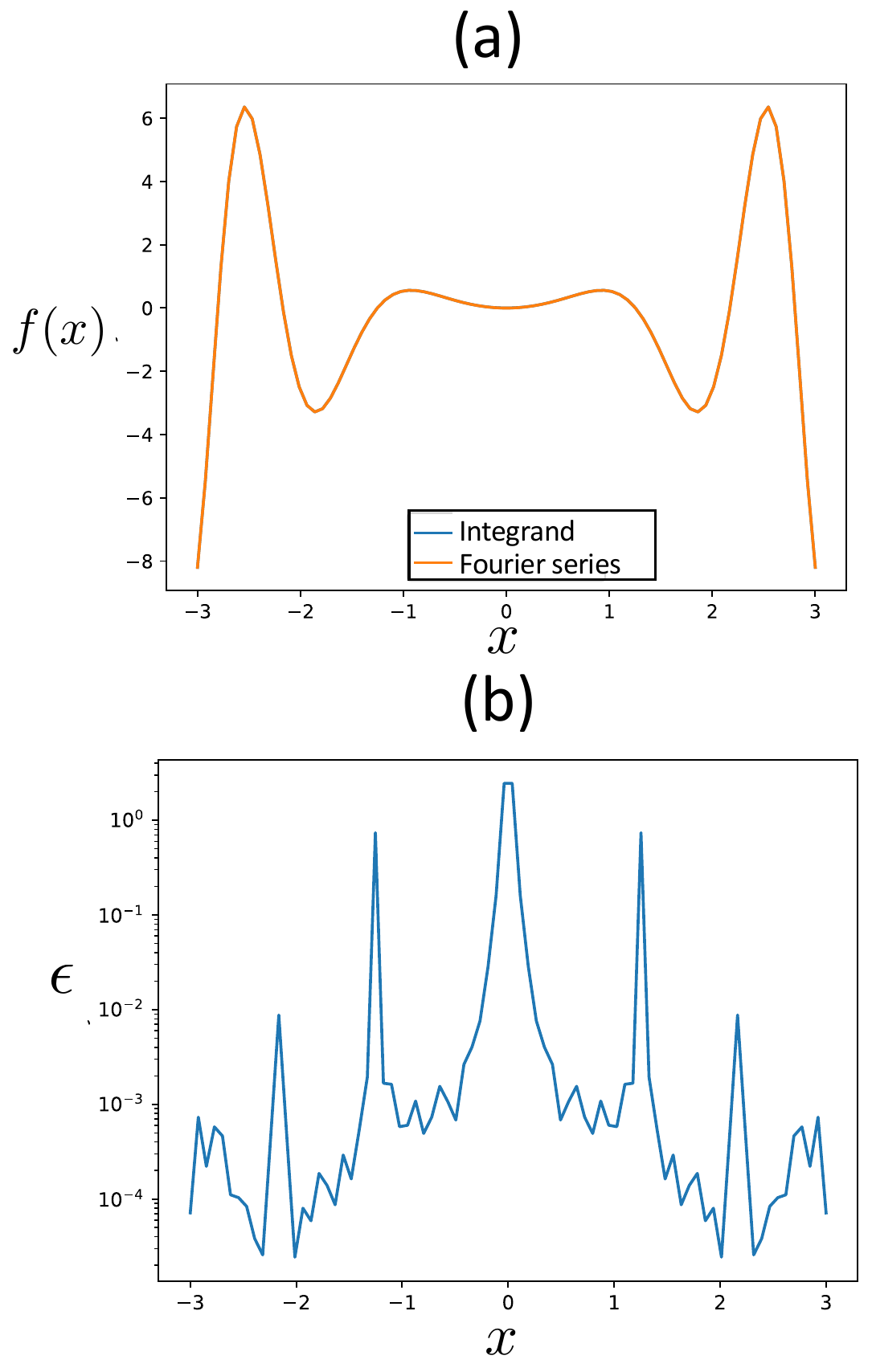}
\caption{[Color online] (a) $f(x) = x^2 \cos\left( x^2 \right)$ and Fourier approximation (almost indistinguishable); (b) relative error between exact and Fourier approximation.}	
\label{fig6}
\end{figure}

\begin{figure}
	\centering
	\includegraphics[width=0.9\linewidth]{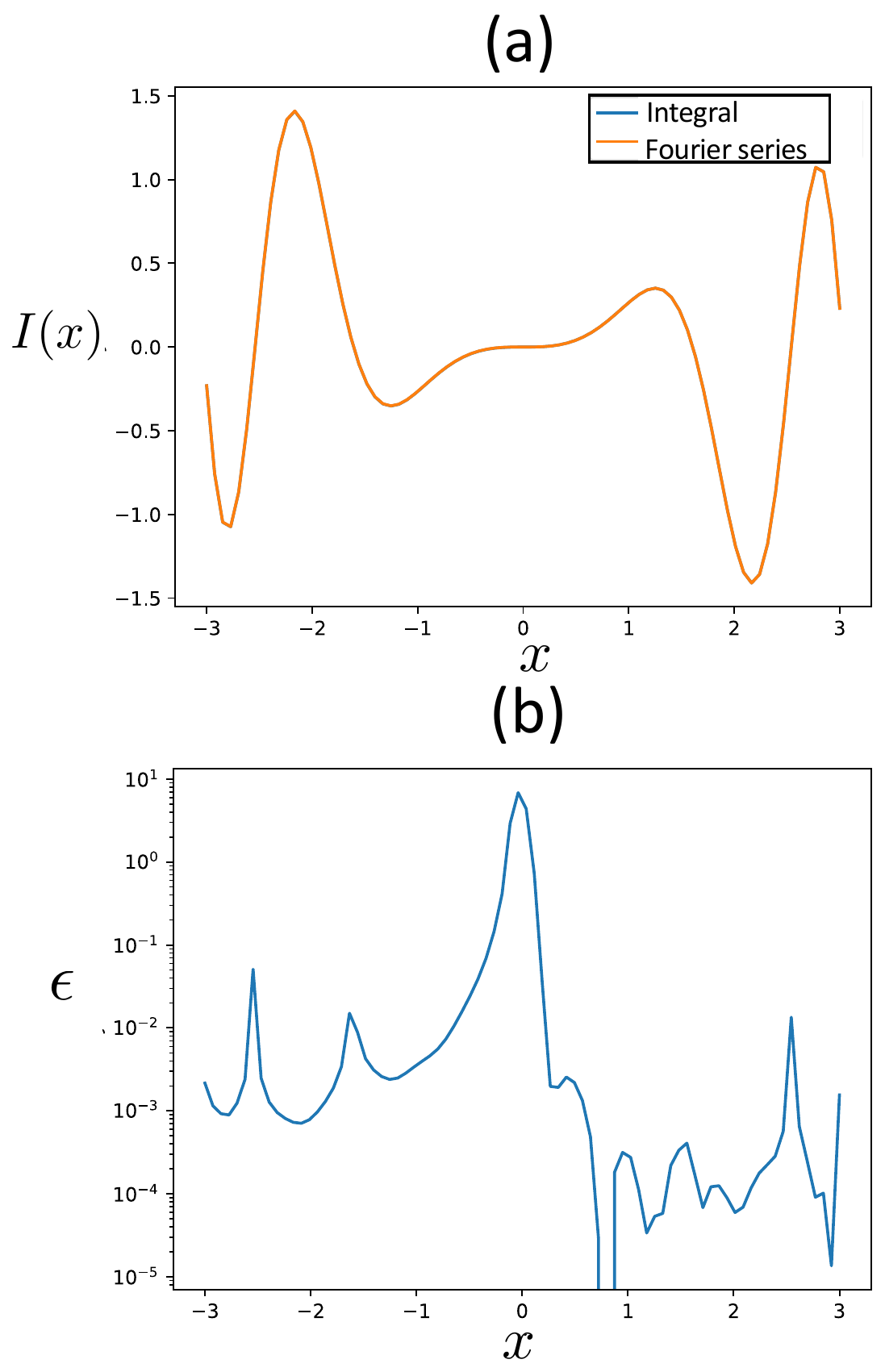}
	\caption{[Color online] (a) Integral of Fig.\ref{fig6}, exact and Fourier approximation (almost indistinguishable); (b) relative error between exact and Fourier approximation.}	\label{fig7}
\end{figure}

In addition, we computed the integral 
\beq 
I(a,b) = \int_a^b x^2 \cos\left(x^2\right) dx. 
\eeq
The result of this integral has actually a closed-form expression, which is
\beq
I(a,b) = \frac{1}{4} \left( 2 x \sin\left( x^2 \right) - \sqrt{2 \pi} S\left(\sqrt{\frac{2}{\pi}} x\right) \right)\Bigg|_a^b,  
\eeq
with $S(z)$ the Fresnel integral 
\beq
S(z) = \int_0^z \sin\left( \pi t^2/2 \right) dt.
\eeq
We have implemented the integral in the interval $[-3, 3]$ with the same parameters as in the previous example, and the results are shown in Fig.\ref{fig6} (integrand and relative error) and Fig.\ref{fig7} (integral and relative error), using 12 qubits, 34 Fourier modes and a mesh of 80 points. Again, in just a few minutes the accuracy in both integral and integral is very high, with very small relative errors, in both cases around $10^{-3}$. 

\begin{figure}
	\centering
	\includegraphics[width=0.9\linewidth]{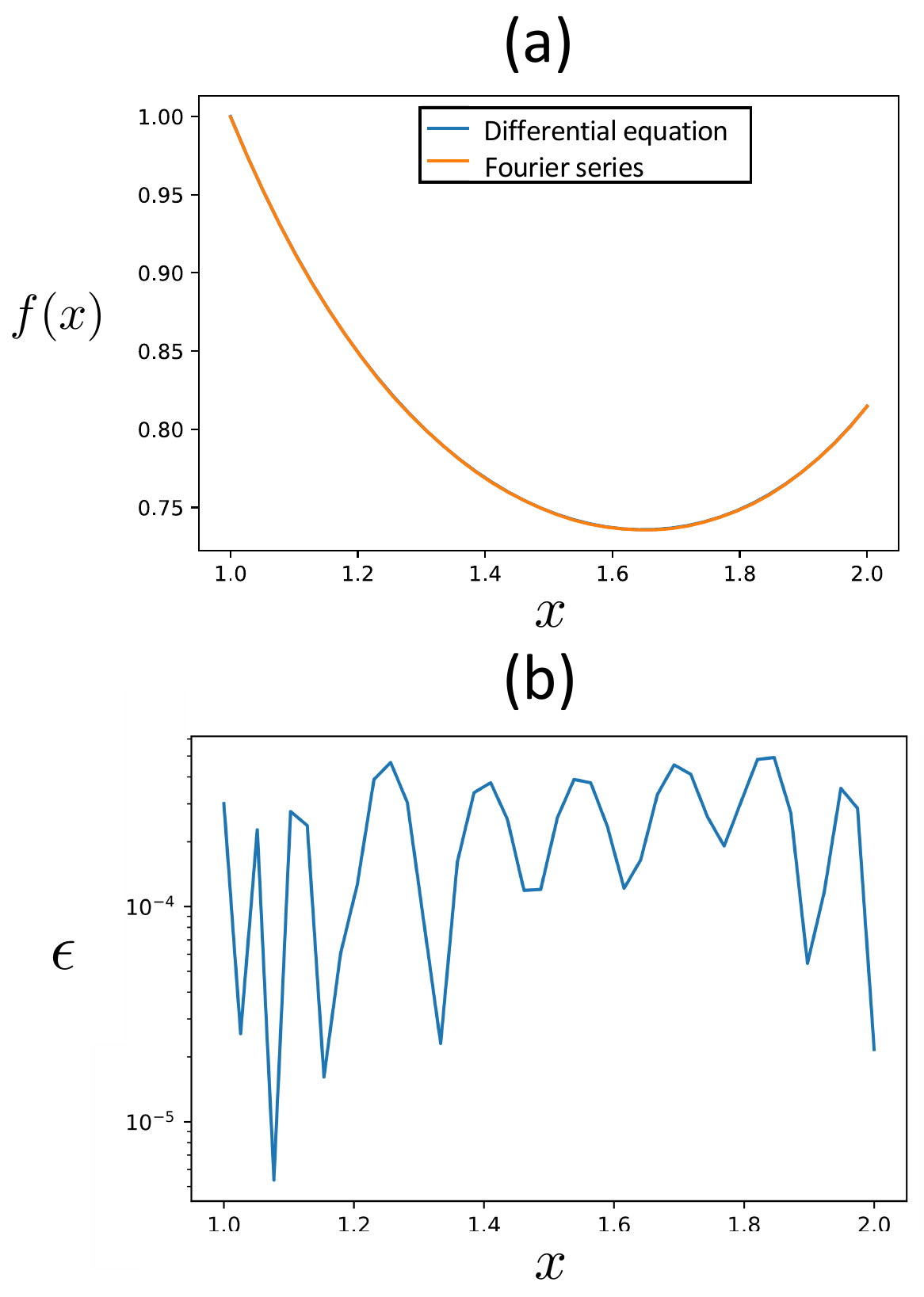}
\caption{[Color online] (a) Exact solution of the differential equation $x f'(x) + f(x) = f^2(x) x^2 \log(x)$ with $f(1) = 1$, and Fourier approximation (almost indistinguishable); (b) relative error between exact and Fourier approximation.}	
	\label{fig8}
\end{figure}

\begin{figure}
	\centering
	\includegraphics[width=0.9\linewidth]{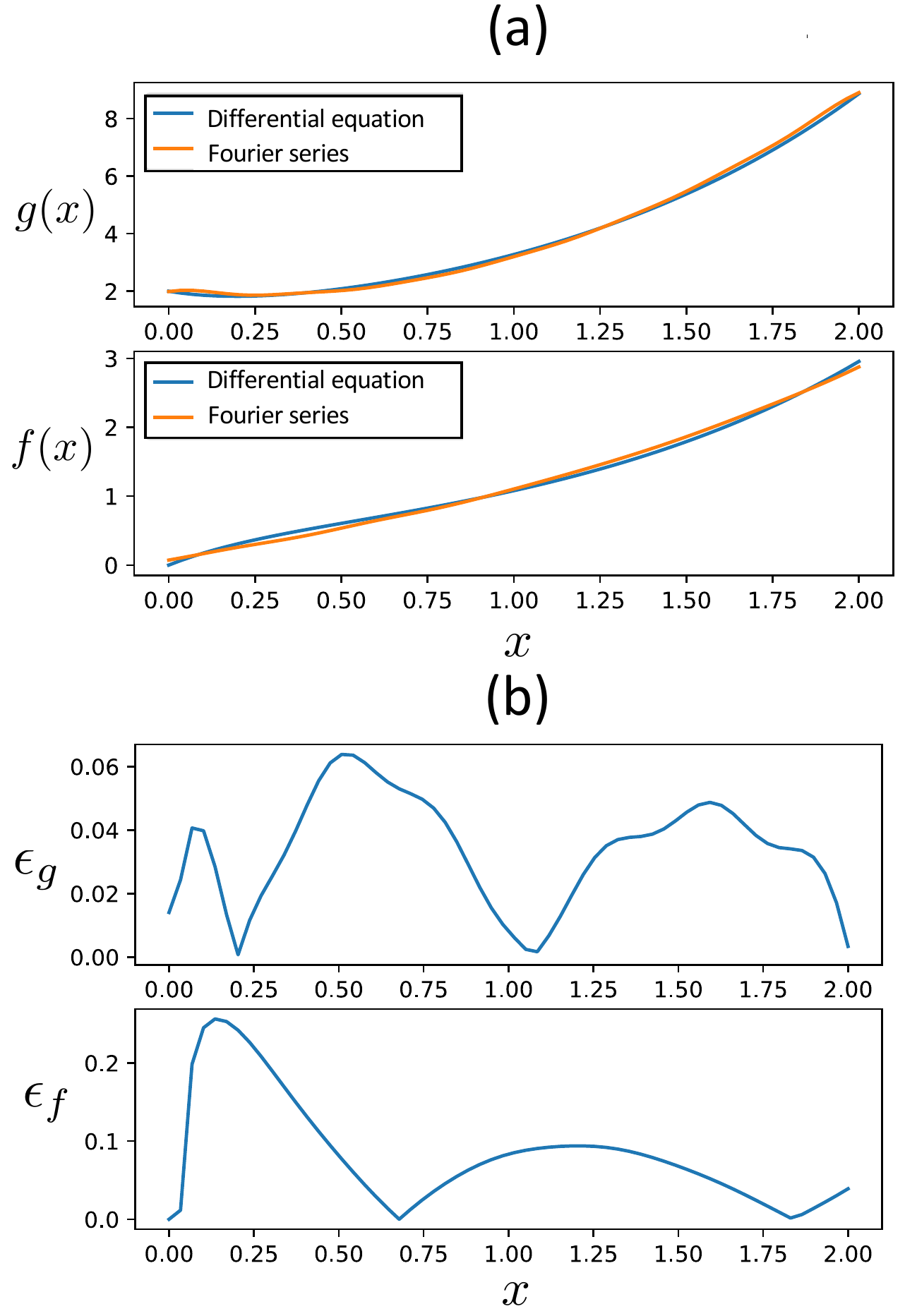}
\caption{[Color online] (a) Exact solutions $g(x), f(x)$ of the system of differential equations in Eq.(\ref{sys}), and their Fourier approximations; (b) relative errors between the exact solutions and their Fourier approximations.}
	\label{fig9}
\end{figure}

\subsection{Differential equations and systems thereof}
As is well known, transforms and series expansions are a way to solve differential equations, as well as systems of such types of equations. Take for instance the case of the linear differential equation 
\beq
a_0(x) f(x) + a_1(x) f'(x) + \cdots + a_n(x) f^{(n)}(x) = b(x),  
\eeq
where $a_0(x), \cdots, a_n(x)$ and $b(x)$ are some functions and $f'(x), \cdots, f^{(n)}(x)$ the derivatives of function $f$ with respect to variable $x$. The solution to this problem is the one that minimizes the functional 
\beq
F[f(x)] \equiv \left(a_0(x) f(x) + \cdots + a_n(x) f^{(n)}(x) - b(x) \right)^2,  
\eeq
with respect to $f$ at every point $x$. Boundary conditions of the solution can always be incorporated as constraints in the minimization problem via Lagrange multipliers. By using the truncated Fourier expansion of $f(x)$ with exponential functions and in a domain of size $2L$, the above functional is approximated as
\beq
F[f(x)] \approx \left(\sum_{l = -K/2}^{K/2} \sum_{\alpha = 0}^n a_\alpha(x) (i 2 \pi l/L)^\alpha c_l e^{i 2 \pi l x / L} - b(x) \right)^2
\eeq
for every point $x$. The differential equation can then be solved by further expanding $a_0(x), \cdots, a_n(x)$ and $b(x)$ into Fourier series. After aggregating all exponential terms in the resulting sum, one ends up with an expansion of the form 
\beq
F[f(x)] \approx \sum_{l = -K/2}^{K/2} g_l(\vec{c}) e^{i 2 \pi l x_\mu / L} , 
\eeq
with $g_l(\vec{c})$ some known function of coefficients $\vec{c}$ (arranged in vector notation). Because of the orthogonality of exponentials, minimizing the above approximated functional over $\vec{c}$ amounts to solve the set of minimization problems 
\beq
\min_{\vec{c}} g_l(\vec{c}) ~~ \forall l,   
\eeq
with the constraints corresponding to boundary conditions as Lagrange multipliers. As a solution of the above set of minimizations we obtain an approximated solution to the original differential equation, in the form of a Fourier expansion. Importantly, this procedure can be generalized without too much effort to more complicated situations, such as multidimensional functions with partial differential equations, vector and tensor fields, non-linear differential equations, systems of partial differential equations, integro-differential equations, and more. 

To validate our approach we have implemented the following calculations. First, we solved the non-linear differential equation 
\beq
x f'(x) + f(x) = f^2(x) x^2 \log(x); ~~~ f(1) = 1, 
\eeq
in the domain $x \in [1, 2]$. The solution can be found in Fig.\ref{fig8}, where we show the exact solution as well as the Fourier approximation using 8 qubits, 11 modes and a mesh of 80 points for the optimization. The method works really well, producing in few minutes a very small relative error around $10^{-4}$. Next, we considered the following coupled system of differential equations: 
\beqa 
g'(x) &=& - g(x) + 6 f(x); ~~~ g(0) = 2, \nonumber \\Ê
f'(x) &=& g(x) - 2 f(x); ~~~~~ f(0) = 0, 
\label{sys}
\eeqa
in the domain $x \in [0, 2]$. The solution can be found in Fig.\ref{fig9}, where we used 12 qubits, 7 modes and a mesh of 60 points for optimization. As we can see in the figures, the method produces in a matter of minutes good approximations of relatively low error in the range $10^{-1} - 10^{-2}$.  

\section{Conclusions and outlook}
\label{sec4}

In this paper we have presented a quantum algorithm to optimize functions of variables with continuous domains. The algorithm is based on a variational quantum circuit, in combination with an encoding of the function's variables in the continuous parameters of the individual qubits. By doing single-qubit quantum state tomography, our procedure allows to optimize functions of up to $3n$ continuous variables using a quantum computer with $n$ qubits. We have implemented classical simulations of the algorithm and arrived to the conclusion that it is remarkably fast and stable, with our simulations being able to optimize complicated functions of 28 variables using 14 qubits (using a pure-state encoding) in just a few seconds. 

Furthermore, we have explained why our quantum optimization algorithm is the building block of a complete toolbox for mathematical analysis with quantum computers. Solving continuous optimization problems allows us to compute, for instance, series expansions, Fourier analysis, and integrals, as well as to solve differential equations and systems thereof. We have provided explicit examples and benchmarks for these calculations. But as a matter of fact, the possibilities are unending, as well as the applications in science and engineering. 

The fact that our optimization algorithm is heuristic makes it also particularly resilient to noise. Additionally, because of being heuristic, it is difficult to do an analysis of the computational complexity of the algorithm, in the same way that one can not say too much about the computational cost of VQE algorithms, and even of classical neural networks. Nevertheless, what is clear is that our algorithm makes use of all the powerful features of quantum computers: entanglement, superpositions, and now continuous encoding. The tomography part could also be implemented, eventually, via quantum methods, increasing even more the efficiency. 

The algorithm and procedures presented here show how a quantum computer with just a bunch of few noisy qubits can already offer value in real complex problems. As an example, using our algorithm, a quantum computer of 127 qubits, such as the largest quantum processor commercially available as of today \cite{ibm}, would be able to do mathematical analysis for 381-dimensional functions of continuous variables, directly in an analog way.  This capabilities are therefore already within reach with present day quantum processors. We believe that the quantum algorithms presented here are candidates to show quantum advantage in the short term, as well as to provide valuable quantum applications in NISQ devices for real-life industry challenges. 

\bigskip
{\bf Acknowledgements.-} The authors acknowledge
DIPC, Ikerbasque, Basque Government and Diputaci\'on
de Gipuzkoa for constant support, as well as insightful discussions with the technical teams from Multiverse
Computing and DIPC on the algorithms and technical implementations. We also acknowledge Elena Zabala from Galbaian IP and Robert Harrison from Sonnenberg Harrison for supporting us with several patent applications in connection with parts of the work discussed here. 

\bibliography{bibliography}

\end{document}